\RequirePackage{fix-cm}
\RequirePackage{amsmath}

\documentclass[twocolumn,epjc3]{svjour3} 

\RequirePackage{amssymb}
\RequirePackage{graphicx}
\RequirePackage{bm}
\RequirePackage{lmodern}
\RequirePackage{verbatim}
\RequirePackage{units}
\RequirePackage{graphicx}
\RequirePackage[utf8]{inputenc}
\RequirePackage[english]{babel}

\journalname{Eur. Phys. J. C}

\begin{document}

\title{Stratified
scalar field theories of gravitation with self-energy term
and effective particle Lagrangian
} 

\author{Diogo P. L. Bragan\c{c}a\thanksref{e1,addr1,addr2}
        \and
        Jos\'e P. S. Lemos\thanksref{e2,addr1} 
}

\thankstext{e1}{e-mail: braganca@stanford.edu}
\thankstext{e2}{e-mail: joselemos@ist.utl.pt}

\institute{Centro de Astrof\'{\i}sica e Gravita\c c\~ao - CENTRA,
Departamento de F\'{\i}sica, Instituto Superior T\'{e}cnico - IST, 
Universidade de Lisboa - UL, Av. Rovisco Pais 1, 1049-001 Lisboa, Portugal \label{addr1}
           \and
           \emph{Present Address:} Stanford Institute of Theoretical Physics - SITP,
           Department of Physics, Stanford University, Stanford, CA 94305, USA \label{addr2}
}

\date{Received: date / Accepted: date}

\maketitle

\begin{abstract}
We construct a general stratified scalar theory of gravitation from
a field equation that accounts for the
self-interaction of the field and a
particle Lagrangian, and calculate its post-Newtonian
parameters. Using this general framework, we analyze several specific
scalar theories of gravitation and check their predictions for the
solar system post-Newtonian effects.
\end{abstract}

\section{Introduction}

Newtonian gravitation is a scalar theory
in which the gravitational interaction is described
by a gravitational scalar field or potential $\Phi$,
that satisfies
a Poisson equation. It is complemented
by Newton's second law of mechanics for the
trajectories of particles moving in the
gravitational potential $\Phi$.

With the advent of
special relativity it became clear that energy and mass
are equivalent and so any form of energy, like the
energy contained in any physical field, produces
gravitational field and also gravitates.
The gravitational field, possessing itself gravitational
energy, should thus produce additional
gravitational field  and thus
the Poisson law should be modified
to contain this field self-energy.
A consistent generalization of Newtonian's gravitation
accounting for the weight of gravitational self-energy
has been performed by several authors, see
e.g.
\cite{peters1981,giulini1997,Frauendiener2011,franklin2015},
see also 
\cite{Feynman2002} for a discussion of gravitational self-energy
terms.

Special relativity also implied that any proposed theory of
gravitation should be relativistic.  The simplest way is to put
Newtonian's gravitation in a relativistic form.  Scalar gravitational
theories were initiated by Nordstr\"om with the gravitational
potential being treated as a scalar field on a Minkowski background
\cite{nordstrom1912} and then modifying it into a scalar theory in a
conformal background \cite{nordstrom1913}, with its full structure
displayed by Einstein and Fokker \cite{einsteinfokker1914} who showed
that it is a covariant scalar theory in a conformally flat space-time,
i.e., gravitational effects can be
seen as a consequence of having a curved metric generated by a scalar
gravitational potential, 
see also the review by Laue \cite{laue1917}.
The idea of conformal
theories of gravitation were resurrected by 
Littlewood \cite{littlewood1953}, whose
theory  arose the interest of Pirani
\cite{pirani1954}, and further developed by
G\"ursey \cite{gursey1953}, 
Bergmann \cite{bergmann1957},
and Dowker \cite{dowker1965}.
Several other studies analyzed their
properties 
\cite{sexl1967,freund1968,sexl1970,deserhalpern1970,linden1971,coleman1971,Shapiro1993,wattmisner1999,calogero2003,calogero2004,ravndal2004,sundrum2004,gls2008,deruelle2011,deruellesasaki2011,romero2012}.
Reviews, analyses, and modifications of these
scalar theories can also be seen
in \cite{Gupta1957,schild1962,wellnersandri1964,harvey1965,withrowmorduch1965,guth1972,Doughtylagrangian1990,norton1992,giulini2008}.
Scalar conformal theories of gravitation are simple, interesting, didactic,
but suffer from the problem that due to the conformal character
and the coupling of electromagnetism with gravitation, they
yield a zero light deflection in the presence of a gravitational field.

An additional interesting set of theories of gravitation
that involve a scalar field 
are the stratified scalar theories. These
theories are not purely scalar, they also
possess
a universal reference frame, where a
universal time $t$, and thus a universal vector field,
is defined with
the space slices composing a stratum that is conformally flat.
Einstein was the first to compose such a theory
in which the velocity of light is a variable
quantity that plays the role of
the gravitational potential \cite{einstein1912}
and was also developed by Abraham
\cite{abraham1912}. 
Other stratified theories were composed
by Papapetrou 
\cite{papapetrou19541,papapetrou19542,papapetrou19543},
Yilmaz \cite{yilmaz1958,yilmaz1962},
Whitrow and Morduch \cite{withrowmorduch1960},
Page and Tupper \cite{pagetupper1968}, Rosen \cite{rosen1971},
Ni \cite{ni1972,ni1973},
and  Broekaert \cite{broekaert2008}.
In Ni \cite{ni1972} a review  of 
stratified scalar theories is given. 
Stratified scalar theories can bypass the 
light deflection problem.
Further, these theories, due to the existence of
a preferred vector field, break Lorentz symmetry
and thus could be candidates to a fundamental theory
where Lorentz symmetry is not essential.

Each of these particular scalar theories, be they
conformal or stratified, 
have a field equation and a particle
Lagrangian that together give the particle trajectories in the external
gravitational field. 
The theories predict specific
values for the classical gravitational tests, see
in addition
\cite{Phipps1986,Peters1986,Phipps1987,Ni2016b,Roseveare1982}.
To compare the different theories in these tests, it should then be enough to compare just a few specific parameters of the field equation and of the particle Lagrangian of each theory.
They also have different post-Newtonian effects,
and we can compare them using the Parametrized Post Newtonian (PPN)
formalism
\cite{will1993,will2014}.
The PPN formalism was devised for confronting
general relativity and other theories
of gravitation with observational data
\cite{will1993,will2014,nordtvedt1972,will1976}.
In order to easily compare theories defined by a field equation and particle Lagrangian, a systematic formalism to get PPN parameters from some parameters of the field equation and of particle Lagrangian would be of great help.
General relativity, so far the
most successful theory,
is a tensorial theory
of gravitation.
But, it is believed 
that both at the quantum gravity level and
at cosmological
scales there are
corrections
to general relativity.
On one hand, these corrections
bring complexity and give rise to
new fields
that enter the scene in the same footing
as the tensor field of general relativity.
Indeed, extensions of general relativity
admitting, in addition to the metric tensor field,
vector and scalar fields,
have been proposed as alternatives
theories of gravitation \cite{will1993,will2014}.
On the other hand, general relativity
could be emergent from some underlying
simpler phenomena,
such as atoms of spacetime, perhaps in the form
of simple scalar or vectorial fields.
Thus,
scalar theories
of the type just mentioned, or even
vectorial theories,
can be sought for,
although vectorial
theories of gravitation modeled 
in Maxwell electromagnetism suffer from
the drawback of admitting negative field energy
not admissible for the gravitational field.

In this paper we propose a general stratified scalar theory of
gravitation postulating first a field equation that accounts for the
self-interaction of the gravitational field and second a Lagrangian for
describing particle motion in the gravitational field.
Conformal scalar theories of gravitation are also obtained as a specific case of this general stratified scalar theory.
We also give a direct method to compute the two PPN
parameters that affect the classical solar system tests, namely,
Mercury's perihelion precession, light
deflection, gravitational redshift, and the Shapiro effect. This
method reads the PPN parameters directly from the field equation and
the particle Lagrangian.
Using specific scalar
theories of gravitation we then confront them
with experimental data. 
This method provides a simple, straightforward way to compare between scalar theories of gravitation.

This paper is organized as follows.  In Sec~\ref{sec:theory}, we
define the general stratified scalar theory by postulating a gravitational
field
equation that accounts for self-interaction and a particle Lagrangian
that gives the particle motion.
In Sec.~\ref{sec:wfl}, we calculate the weak field limit and two
post-Newtonian parameters of the theory.  In
Sec.~\ref{sec:application}, we use the general scalar theory to compute
PPN parameters for specific scalar theories of gravitation.
In Sec.~\ref{sec:conc} we conclude.

\section{Building a general stratified scalar field
theory of gravitation with self-interaction} \label{sec:theory}

\subsection{Postulates and equations of the theory}

In a
general stratified scalar gravitational theory
it is necessary to define an existing prior background
structure \cite{ni1972}
(see also \cite{will1993}).
Moreover, to build a gravitational theory,
one needs to know how
a particle moves in a gravitational field
and how 
gravitation is generated by matter.
We thus have to  (i) define
an existing prior background
structure, (ii) give a general
field equation, and (iii) give
a particle Lagrangian for the
particle's trajectories.

The stratified scalar gravitational theory
we are going to work with 
is defined on a background Minkowski spacetime, with line element $ds_{\rm M}$ given by 
\begin{align}
\label{eq:minkowski}
ds_{\rm M}^2= \gamma_{ab} dx^a dx^b,
\end{align}
where $\gamma_{ab}$ is the flat spacetime metric, and
$a$ and $b$ run from 0 to 3. Note that the coordinates
defining Eq.~(\ref{eq:minkowski}) need
not be Minkowskian coordinates and so in general
$\gamma_{ab}$ need not be $\eta_{ab}={\rm diag}\,(-1,1,1,1)$.
We stratify the theory using a universal time parameter $t$. This
parameter is a scalar field satisfying the following equations
\begin{align}
\nabla_b \nabla_a t &=0\,, \\
(\nabla_a t) (\nabla_b t) \, \gamma^{ab} &= -1\,,
\end{align}
where $\nabla_a$ is the covariant derivative with respect to the metric $\gamma_{ab}$.

To define the field equation for the gravitational scalar
field 
$\Phi$, we generalize the Poisson equation in order to account for a self-interaction of the gravitational field. We then assume that the field equation is given by
\begin{align}
\label{eq:poisson}
\square \Phi = 4 \pi G \rho - k \frac{\gamma^{ab}\nabla_a \Phi \nabla_b \Phi}{c^2} \,,
\end{align}
where $\square$ is the d'Alembertian, $\nabla_a$ is the covariant derivative, both  with respect to the metric $\gamma_{ab}$, 
$G$ is Newton's gravitational constant, $\rho$ is the gravitational source density
and $k$ is a dimensionless constant.

The gravitational source density $\rho$ is a scalar and thus can be defined in two different ways, namely,
$\rho = -T_{ab} u^a u^b$ where $u^a$ is the four-velocity of the source
with respect to $\gamma_{ab}$, or
$\rho = -\gamma_{ab} T^{ab}$. 
Although in our work we do not need to specify $\rho$, it is important to note that the Kreuzer experiment is not compatible with the first possibility $\rho = -T_{ab} u^a u^b$ \cite{ni1972,will1976}.
This entails that $\rho = -\gamma_{ab} T^{ab}$ is the most realistic choice to be used in Eq.~\eqref{eq:poisson}. 
In the case of electromagnetic radiation, we then have $\rho = 0$, which means that light does not generate gravitational field. 
As we shall see, this does not necessarily imply that light is not bent by gravity; this is only the case for conformal scalar theories of gravitation.

Note also that we explicitly include a self-interaction term
with coupling constant $k$, but even though different scalar theories of gravity yield a particular $k$, we will let $k$ have a priori any value.
See below a
derivation of
the modified Poisson equation Eq.~(\ref{eq:poisson}).

We now formulate how test particles behave in the theory. 
For this, we
impose that the particle's trajectories are the geodesics of 
a metric $g_{ab}$ which itself is generated by the scalar gravitational potential $\Phi$
and the universal time $t$. 
Thus, we write quite generally \cite{ni1972}
\begin{align}
ds^2 &= g_{ab} dx^a dx^b  \nonumber \\
&= - (g(\Phi) - f(\Phi)) c^2 dt^2 + f(\Phi) \gamma_{ab} dx^a dx^b \,,
\label{eq:physmetric}
\end{align}
where $c$ is the velocity of light, 
$dt$ is the differential of the universal time $t$, $\Phi$ is the gravitational potential, and $g$ and $f$ are two scalar functions of $\Phi$.
Since $g_{ab}$ in Eq.~(\ref{eq:physmetric}) defines the geodesics it must be considered
the physical metric.
In general, this physical metric breaks Lorentz symmetries (even though these are preserved in the background metric). This Lorentz symmetry breaking arises also in standard cosmology (e.g. the Cosmic Microwave Background indicates a preferred set of reference frames), and that may indicate that at a fundamental level there must be a breaking of Lorentz symmetries, and as such these stratified theories should not be discarded a priori.
Note that light propagates in the null geodesics of this metric, and therefore, in the general case, does not follow straight lines. 
For the physical metric, the action $S$ for a particle trajectory is then
\begin{align}
S = - m c^2 \int d\tau \,,
\label{eq:action}
\end{align}
where $d\tau=\frac1c\sqrt{-g_{ab} dx^a dx^b}$, and $m$ is the mass of the particle.
From the standard definition of a
Lagrangian $L$, namely,
\begin{align}
S = \int L \, dx^0 \,,
\label{eq:lagdef}
\end{align}
in a given coordinate system $x^a$,
with $x^0$ being some time coordinate, 
we get, from Eqs.~\eqref{eq:physmetric}, \eqref{eq:action}, and \eqref{eq:lagdef}, the following particle Lagrangian,
\begin{align}
\label{eq:genlag}
&L = \\
& - mc\sqrt{(g(\Phi) - f(\Phi)) c^2 \left(\frac{dt}{dx^0}\right)^2 - f(\Phi) \gamma_{ab} \frac{dx^a}{dx^0} \frac{dx^b}{dx^0}} \,. \nonumber
\end{align}
In a stratified theory, without loss of generality,
it is clearly convenient, from 
\eqref{eq:genlag}, that $x^0$ should be identified with $t$,
 $x^0=t$. In this case there are further simplifications, namely,
$\gamma_{00} =-c^2$, and imposing further that the metric is static
one can set
$\gamma_{0\alpha}=0$, where $\alpha$ is a spatial index
running from 1 to 3. Using these facts, the Lagrangian \eqref{eq:genlag} becomes
\begin{align}
L = -mc \sqrt{g(\Phi) c^2 - f(\Phi) v^2}\,,
\label{eq:partlag}
\end{align}
where the particle's 3-velocity $v$ is defined through
the relation
\begin{align}
v^2 = \gamma_{\alpha \beta} \frac{dx^\alpha}{dx^0} \frac{dx^\beta}{dx^0} \,.
\end{align}
Since in special relativity the Lagrangian for a particle is \cite{will1993}
$L = -mc \sqrt{ c^2 -  v^2}$,
we see that
to get the correct special relativity limit,
we have to impose the following condition on the functions
$f$ and $g$, $f(0)=g(0)=1$, where
without loss of generality we are assuming that no gravitational field
means $\Phi=0$ rather than $\Phi={\rm constant}$.
See below 
a derivation of the particle Lagrangian given in
Eq.~(\ref{eq:partlag}).

It is important to remark that this formalism can be used to study
conformally flat theories as well. In fact, to cancel the influence of
the universal time $t$, it is sufficient to set $g=f$ in
Eq.~\eqref{eq:genlag}. Choosing a reference frame in which
$\gamma_{0\alpha}=0$, the Lagrangian becomes
\begin{align}
L = -mc \sqrt{f(\Phi)}\sqrt{c^2 - v^2}\,,
\label{eq:partlag2}
\end{align}
which is simply Eq.~\eqref{eq:partlag} with $g=f$, as expected,
i.e., it represents a conformal scalar field theory. We will use this fact throughout the paper. 
Finally, note that, in this case, Eq.~\eqref{eq:physmetric} shows that light propagates in straight lines.

\subsection{Derivation of the Poisson equation
with a self-energy term and of the particle Lagrangian}

\subsubsection{Derivation of the Poisson equation
with a self-energy term}
\label{motiv1}
We can motivate our field equation~(\ref{eq:poisson}) through the following scheme. 
In electrodynamics, the electromagnetic energy is
stored in the field with a positive energy density $\rho_{\rm EM}$ given by
$
\rho_{\rm EM}=\frac{1}{2}\left(
| \boldsymbol{E} |^2 +  | \boldsymbol{B}|^2\right)
$, where $\boldsymbol{E}$ and $\boldsymbol{B}$
are the electric and magnetic fields, respectively.
A similar expression can be obtained
for the gravitational field in Newtonian gravitation.
Indeed, using Poisson's equation,
\begin{align}
\nabla^2 \Phi = 4 \pi G \rho\,,
\label{egrav0}
\end{align}
one can show that the total gravitational potential energy $E_{\rm
grav}$ in a given volume $V$ can be written as (see
e.g. \cite{peters1981})
$
E_{\rm grav} = \frac{1}{2} \int_V \rho \Phi \, \mathrm{d} V
= \int_V -\frac{\vert \nabla \Phi \vert ^2}{8 \pi G} \mathrm{d} V
$,
where $\rho$ is the matter density, $\Phi$ is the gravitational
potential, $G$ is Newton's gravitational constant, $\nabla$ is the
gradient operator, and an integration by parts has been performed.
Therefore, in the Newtonian theory of
gravitation, one may define a gravitational field energy density as
\begin{align}
\rho_{\rm grav} \equiv -\frac{\vert \nabla \Phi \vert^2}{8 \pi G}\,.
\label{egrav}
\end{align}
It is interesting to note that this is a negative definite energy
density. This stems from the fact that gravity in the Newtonian theory
is exclusively an attractive force.
 
Now, we assume
that the energy of the gravitational field can also gravitate.
Thus, Eq.~(\ref{egrav0}) with Eq.~(\ref{egrav}) yields
$\nabla^2 \Phi = 4 \pi G \rho-\frac12 \vert \nabla \Phi \vert^2$.
Notice that this approach is not self-consistent (for self-consistent constructions see \cite{franklin2015,freund1968}); nonetheless, this modified Poisson equation is valid to the first post-Newtonian order, which is the order we are interested in for PPN formalism purposes.
Moreover, to put in a relativistic setting, and in order to get a Lorentz scalar, we
replace
the Laplacian in Poisson's equation~(\ref{egrav0}) by the d'Alembertian
and generalize the gradient $\nabla$ to the covariant derivative $\nabla_a$.
Then we get 
$\square \Phi = 4 \pi G \rho - \frac12
\frac{\gamma^{ab}\nabla_a \Phi \nabla_b \Phi}{c^2}$,
where $\gamma_{ab}$ is the Minkowski metric.
In order to consider a more general theory, we let the factor $\frac12$ that multiplies
$\frac{\gamma^{ab}\nabla_a \Phi \nabla_b \Phi}{c^2}$
be undetermined, call it $k$,
obtaining thus the
sought for equation
\begin{align}
\label{eq:poissonrep}
\square \Phi = 4 \pi G \rho - k \frac{\gamma^{ab}\nabla_a \Phi \nabla_b \Phi}{c^2} \,,
\end{align}
i.e., Eq.~\eqref{eq:poisson}.

\subsubsection{Derivation of the particle Lagrangian}
\label{motiv2}

We can motivate the definition of our particle Lagrangian~(\ref{eq:partlag}) 
from an expression for the energy of a particle in a curved static spacetime.

A static spacetime with metric $g_{ab}$ as
can be written 
\begin{align}
\label{eq:metric}
ds^2 = g_{ab}\, dx^a dx^b=
g_{00} dt^2 + g_{\alpha \beta} \, dx^\alpha dx^\beta \,,
\end{align} 
where Latin indices $a,b$ run from 0 to 3,
Greek indices $\alpha,\beta$ correspond to the spatial part 
of the metric and run from 1 to 3,
$g_{00}=g_{00}(x^\alpha)$,
and in general $g_{\alpha \beta}=g_{\alpha \beta}(x^\alpha)$.
We assume asymptotic flatness.

We now calculate the appropriate expression for the observed energy $E$
of a particle 
measured by a static observer in the metric given by Eq.~\eqref{eq:metric}.
The four-velocity $v^a$ of the particle is 
$
v^a = \frac{dx^a}{d\tau} 
$,
where $\tau$ is the proper time of the particle,
and its four-momentum $p_a$ is
$
p_a = m \, g_{ab} v^b
$.
If an observer has four-velocity $u^b$, say,
then the total observed energy of the particle measured in the local observer's  frame is
$
E = - p_b u^b
$. Identifying the particle's energy $E$ with the Hamiltonian $H$, we can then
find the  expression for the Lagrangian $L$.

To start we assume that $g_{\alpha \beta}(x^\alpha)$
is a flat spatial metric, not necessarily Euclidean,
afterwards we will relax this assumption. Thus, calling
$\gamma_{\alpha\beta}$ a general flat 
spatial metric, we put
\begin{align}
g_{\alpha \beta} = \gamma_{\alpha \beta}\,.
\label{gabeta}
\end{align}
Consider now the metric given in Eq.~\eqref{eq:metric}
and consider a coordinate system in which the
metric~(\ref{gabeta}) is 
diagonal, for instance in static
spherical coordinates. Then, we have that the 
particle's four velocity can be written as
$
v^a = \frac{dx^a}{dt} \frac{dt}{d\tau} =  \frac{dx^a}{dt} v^0 
$,
where $v^0=\frac{dt}{d\tau}$.
The spatial components of the velocity $\boldsymbol{v}^\alpha$ defined with respect to the
time $t$ are $\boldsymbol{v}^\alpha=\frac{dx^\alpha}{dt}$
and the square of the spatial velocity is $
\boldsymbol{v}^2= \gamma_{\alpha \beta} \boldsymbol{v}^\alpha \boldsymbol{v}^\beta
$.
Then, since $g_{ab} v^a v^b=-c^2$, we get
$
g_{ab} v^a v^b = g_{00} \left( v^0 \right)^2 + \left( v^0 \right)^2 \boldsymbol{v}^2=-c^2
$.
We can then solve this equation for $v^0$, getting 
$
v^0 = c \, \sqrt{\frac{-1}{g_{00} + \boldsymbol{v}^2}} 
$.
We want the particle energy measured by an inertial static observer
at infinity. 
Such a static  observer has spatial velocity zero, $\boldsymbol{u}^\alpha=0$.
Since such an observer is also a test particle the
above derivation for $v^0$ holds, but now we have to replace
$v^0$ by $u^0$ and $\boldsymbol{v}^\alpha$ by $\boldsymbol{u}^\alpha=0$, an so
$\boldsymbol{u}^2=0$, and from asymptotically flatness $g_{00}=-c^2$.
So,
$
u^0 = c \, \sqrt{\frac{-1}{g_{00}}} = 1
$.
Since the metric is diagonal and $\boldsymbol{u}^\alpha=0$, 
the energy expression
$E = - p_b u^b$
simplifies to
$
E = -m g_{00} \, v^0 u^0 
$.
Then using the expressions for $v^0$ and $u^0$ just found 
we find for the  energy the expression
\begin{align} 
E = - m \,g_{00}\, c\, \sqrt{\frac{-1}{g_{00} + \boldsymbol{v}^2}}  \,.
\end{align}
We proceed by identifying this energy with the Hamiltonian $H$ of the particle,
i.e., $H=E$. 
Writing the Hamiltonian as a function of the spatial components of the momentum of the particle $p_\alpha=m g_{\alpha b} v^b$, we obtain
\begin{align}
H = \sqrt{-g_{00}\, p^2 -g_{00}\, m^2 c^2}\,,
\end{align}
where here $p^2 = \gamma^{\alpha\beta} p_\alpha p_\beta$.
This reduces to special relativity when $g_{00}=-c^2$.
Now, we want the corresponding Lagrangian $L$. 
Using the Legendre transformation relating $H$ and $L$, namely,
$L = \frac{dx^\alpha}{dt} p_\alpha - H$, together with the Hamilton equation $\frac{dx^\alpha}{dt} = \frac{\partial H}{\partial p_\alpha}$,
we can verify that the particle Lagrangian is
\begin{align}
L = -mc \sqrt{-g_{00} - \boldsymbol{v}^2}\,.
\label{lagflat}
\end{align}

We generalize this approach for a
non-flat static metric
whose spatial part $g_{\alpha \beta}$
can be put
in isotropic coordinates. We then write 
\begin{align}
g_{\alpha \beta} = f \, \gamma_{\alpha \beta}\,,
\end{align}
where $\gamma_{\alpha \beta}$ is a flat metric, not necessarily
Euclidean metric, and $f$ is a conformal factor.
With this spatial metric, the particle
Lagrangian given in Eq.~(\ref{lagflat})
becomes
\begin{align}
\label{eq:lagrangian}
L = -mc \sqrt{-g_{00} - f v^2 } \,,
\end{align}
where $v^2 = \gamma_{\alpha \beta} v^\alpha v^\beta$.
Writing $-g_{00} \equiv g(\Phi) c^2$ and $f=f(\Phi)$,
for some $\Phi$ and identifying this $\Phi$ with the
gravitational potential,
we see that the particle Lagrangian in Eq.~(\ref{eq:lagrangian})
is the same as
\begin{align}
L = -mc \sqrt{g(\Phi) c^2 - f(\Phi) v^2}\,,
\label{eq:partlagrep}
\end{align}
i.e., 
we recover the
postulated
particle Lagrangian given in Eq.~\eqref{eq:partlag}.

It is interesting to notice that this particle
Lagrangian Eq.~\eqref{eq:partlagrep}, or Eq.~\eqref{eq:partlag},
is the same as the one that comes from the requirement that
the trajectory in spacetime is a geodesic of the metric $g_{ab}$.
In fact, the particle action in general relativity  $S_{\rm GR}$
is given by
$S_{\rm particle\;GR} = - mc \int d\tau$,
which corresponds to the Lagrangian
$L_{\rm particle\;GR} = -mc \frac{d\tau}{dt}$.
It is immediate to verify that $L_{\rm particle\;GR}=L$.
Therefore, the particle trajectory in our theory is a geodesic of the metric 
$ds^2= -g(\Phi) c^2 dt^2 + f(\Phi)\gamma_{\alpha \beta} dx^\alpha dx^\beta$,
$\gamma_{\alpha \beta}$ being the spatial flat metric.
However, we have derived the Lagrangian Eq.~\eqref{eq:partlagrep}, or
Eq.~\eqref{eq:partlag},
without the requirement that a particle follows a geodesic of
spacetime and without requiring the covariant divergence of the
particle's energy-momentum tensor to vanish. This result is certainly
very interesting.

\section{Weak field limit of the theory} \label{sec:wfl}

\subsection{Choices of $g$ and $f$}

In order to compare the stratified  scalar theory
given in Eqs.~(\ref{eq:poisson}) and~(\ref{eq:partlag})
with experiment, we take advantage of the Parametrized Post Newtonian (PPN) formalism. To use it, we Taylor expand $f$ and $g$ to second order in $1/c^2$. We define second order terms as having the following factors, $\frac{v^4}{c^4}$, $\frac{\Phi v^2}{c^4}$, or $\frac{\Phi^2}{c^4}$.
The function $g(\Phi)$ is expanded as
\begin{align}
\label{eq:g}
g(\Phi)=1 + 2A\frac{\Phi}{c^2} + 2B\frac{\Phi^2}{c^4}  \,,
\end{align}
where $A$ and $B$ are dimensionless constants.
Since $f(\Phi)$ appears already multiplied by $v^2$
in Eq.~\eqref{eq:partlag},
it suffices to expand to order $\Phi/c^2$, so we write
\begin{align}
\label{eq:f}
f(\Phi) = 1 - 2C\frac{\Phi}{c^2}\,,
\end{align}
where $C$ is a dimensionless constant.
To have the correct Newtonian limit in the Lagrangian~(\ref{eq:partlag}), we must have $A=1$. From now on, we therefore consider $A=1$.

In order to further understand the physical meaning of  $B$
and $C$, we interpret the
Lagrangian Eq.~(\ref{eq:partlag})
as containing an interaction term.
For that
we should expand the particle Lagrangian given in Eq.~\eqref{eq:partlag} to second order and analyze how these parameters affect the interaction part of the Lagrangian. We assume that to second order approximation, we can write our Lagrangian as the sum of a free special relativistic part $L_{\rm  particle \;SR}$  plus an interaction part
$L_{\rm int}$,
$L = L_{\rm particle \;SR} + L_{\rm int} = -mc^2 \sqrt{1-\frac{v^2}{c^2}} + L_{\rm int} $.
Expanding the special relativistic term 
to second order, we get
$
L = -mc^2 + \frac12 m v^2 + \frac18 m \frac{v^4}{c^2} +  L_{\rm int} \,.
$
On the other hand, expanding the particle Lagrangian given by Eq.~\eqref{eq:partlag}, we get
$
L = -mc^2 + \frac12 m v^2 \left[ 1 - (2 C + 1) \frac{\Phi}{c^2} \right] + \frac18 m \frac{v^4}{c^2} -
 m \left[ \Phi + \left(B-\frac{1}{2}\right) \frac{\Phi^2}{c^2} \right]
 $.
Comparing these two equations for $L$
gives the interaction Lagrangian to the relevant order,
$
L_{\rm int} = L - L_{\rm SR}~\nonumber
= - m \left[ \Phi + \left(B-\frac{1}{2}\right) \frac{\Phi^2}{c^2} \right] -
\frac12 m v^2 (2 C + 1) \frac{\Phi}{c^2}$.
Analyzing this equation, we can say first that
$B$ affects the effective gravitational mass of the particle.
This is in line with Nordström's interpretation that gravitational mass is affected by the gravitational field itself \cite{nordstrom1913}.
Second, that
$C$ affects the interaction between the particle's kinetic energy and the gravitational field.

\subsection{Spherical symmetric solution of the field equation}

We have to solve the field equation to obtain the gravitational potential to second order.
We solve Eq.~\eqref{eq:poisson} in a spherically symmetric vacuum, i.e., $\rho=0$, using spherical coordinates $(r,\theta,\phi)$, obtaining
$\frac{d\Phi}{dr} = \frac{GM}{r^2 +\frac{kGMr}{c^2}}$,
where $M$ is the gravitational mass of the central body. Up to order $1/c^2$ this yields
\begin{align}
\frac{d\Phi}{dr} = \frac{GM}{r^2} + k \frac{(GM)^2}{c^2 r^3} \,.
\label{dphidr}
\end{align}
Integrating Eq.~(\ref{dphidr}) we obtain
This corresponds to a potential
\begin{align}
\label{eq:potential}
\Phi = -\frac{GM}{r} - k \frac{(GM)^2}{2 c^2 r^2}\,,
\end{align}
with the condition that at infinity $\Phi=0$.

\subsection{PPN parameters}

We may now substitute Eq.~\eqref{eq:potential} into Eqs.~\eqref{eq:g} and~\eqref{eq:f} to get
\begin{align}
g(\Phi) &= 1 - 2 \frac{GM}{c^2 r} + (2B - k) \frac{(GM)^2}{c^4 r^2} \,,\\
f(\Phi) &= 1 + 2 C \frac{GM}{c^2 r} \,,
\end{align}
where we kept only the terms to the order previously mentioned.

We want to verify how the parameters $B$, $C$, and $k$ are related with PPN parameters $\beta$ and $\gamma$ defined in the standard PPN metric on a static, spherically symmetric spacetime in isotropic coordinates by
\begin{align}
\label{eq:PPN}
ds^2 = - &\left( c^2 - 2 \frac{GM}{r} + 2 \beta \frac{(GM)^2}{c^2 r^2} \right) dt^2 +  \nonumber \\
 & \qquad + \left(1 + 2 \gamma \frac{GM}{c^2 r} \right) \gamma_{\alpha \beta} dx^\alpha dx^\beta \,,
\end{align}
where $\gamma_{\alpha \beta}$ is a flat 3-metric.

Using the correspondence between $g_{00}$ and $f$, and $g(\Phi)$ and $f(\Phi)$ and comparing with the standard PPN metric Eq.~\eqref{eq:PPN}, we immediately identify the following relations
\begin{align}
A&=1\,, \nonumber \\
2B - k &= 2 \beta \,, \nonumber \\
C &= \gamma\,.
\label{eq:system}
\end{align}
To be compatible with every solar system test, it is enough to replace $\beta$ and $\gamma$ by the values of general relativity, that is $\beta=\gamma=1$, and guarantee that our theory satisfies the system \eqref{eq:system}.

Therefore, a stratified scalar field theory of gravitation that has
\begin{itemize}
\item A field equation given in this approximation by Eq.~\eqref{eq:poisson},
\item A particle Lagrangian given in this approximation by Eq.~\eqref{eq:partlag},
\item Parameters that satisfy Eqs.~\eqref{eq:system} for $\beta=\gamma=1$,
\end{itemize}
predicts correctly every solar system effect predicted by general relativity. 

Moreover, a conformally flat scalar theory of gravitation that has a field equation given in this approximation by Eq.~\eqref{eq:poisson}, and parameters $A$, $B$, $C$, and $k$, that satisfy Eqs.~\eqref{eq:system} for $\beta=\gamma=1$ would also correctly predict every relativistic solar system effect.

This approach also provides a clean and fast way to compute the PPN parameters $\beta$ and $\gamma$ for most
theories of gravity, whether stratified scalar field theories or conformally flat scalar theories with only one metric potential.

In the next section, we study consistent theories of gravitation with only one metric potential and calculate for each theory the parameters $B$, $C$ and $k$ to verify if they predict the correct solar system effects.

\section{Application to scalar field theories of gravitation:
Stratified and conformal theories} \label{sec:application}

\subsection{Page and Tupper theory}

The Page and Tupper
theory \cite{pagetupper1968} is a stratified scalar field theory of gravitation
and as such is an instance of the set of equations given
in Eqs.~\eqref{eq:poisson}, \eqref{eq:physmetric}, and 
\eqref{eq:partlag}.
In the preferred reference frame, the theory
has the following field equation and particle Lagrangian
\begin{align}
\square \Phi &= 4 \pi G F(\Phi/c_*^2)^4 \rho \label{eq:ptfield}\,, \\
L &= -m c \sqrt{F^2(\Phi/c_*^2) c_*(\Phi)^2 - F^2(\Phi/c_*^2) v^2}\,,
\label{eq:ptlag}
\end{align}
respectively, where $\square$ is the d'Alembertian in the Minkowski metric, $\rho$ is the gravitational source density, e.g. $\rho = T_{ab} u^a u^b$ where $u^a$ is the four-velocity of the source, or $\rho = T^a_a$, and the functions $F(\Phi/c_*^2)$ and $c_*$ are given by
\begin{align}
F &= 1 - a_1 \frac{\Phi}{c^2} + \left(a_1 Q + a_2\right) \left(\frac{\Phi}{c^2}\right)^2 \,,\\
c_*^{2} &= \frac{c^{2}}{1 - Q \frac{\Phi}{c^2} + R \left(\frac{\Phi}{c^2}\right)^2}\,, 
\end{align}
where $a_1$, $a_2$, $Q$, and $R$, are dimensionless constant parameters
in the theory. Here, $c_*$ is interpreted as a variable speed of light.
Using Eqs.~\eqref{eq:physmetric}, \eqref{eq:partlag}, and~\eqref{eq:ptlag}, it is immediate to recover the physical metric of the Page and Tupper theory,
and we get for the line element
$
ds^2 = F^2(\Phi/c_*^2)(-c_*^2(\Phi) dt^2 + dx^2 + dy^2 + dz^2)
$.

The field equation Eq.~\eqref{eq:ptfield} of the Page and Tupper
theory corresponds to $k=0$ in  Eq.~\eqref{eq:poisson}.
The Lagrangian in Eq.~\eqref{eq:ptlag} of the Page and Tupper
theory
corresponds to $g(\Phi)c^2 = F^2(\Phi/c_*^2) c_*(\Phi)^2$ and to $f(\Phi) = F^2(\Phi/c_*^2)$
in  Eq.~\eqref{eq:partlag}.
Expanding to the post-Newtonian order, we get
for $g(\Phi)$ and $f(\Phi)$
\begin{align}
g(\Phi) =& c^2 \left[1 + (Q-2a_1)  \frac{\Phi}{c^2} + \nonumber \right. \\
& \left. \qquad + (a_1^2 + 2 a_2 + Q^2-R) \left(\frac{\Phi}{c^2}\right)^2\right]\,, \\
f(\Phi) =& 1 - 2 a_1 \frac{\Phi}{c^2} \,,
\end{align}
respectively.
Looking at Eqs.~\eqref{eq:g} and~\eqref{eq:f} 
this corresponds to
\begin{align}
A &= \frac12 Q - a_1\,, \\
B &= \frac12 (a_1^2 + Q^2-R) +  a_2 \,,\\
C &= a_1 \,.
\end{align}
In order to have the correct Newtonian limit, we must have $A=1$, that is $Q = 2 a_1 + 2$.
Therefore, the PPN parameters defined
in
Eq.~\eqref{eq:PPN}, obeying the relation in 
Eq.~\eqref{eq:system}, 
are given by
\begin{align}
\beta &= \frac12 (a_1^2 + Q^2-R) +  a_2 \nonumber
\label{beta2}\\
&= \frac12 (5 a_1^2  + 8 a_1 + 4 -R) +a_2\,, \\
\gamma &= a_1 \label{a12}
\,.
\end{align}
The PPN parameter
$\beta$ has to be one to account for the solar system tests.
We see from 
Eq.~\eqref{beta2}
that the Page and Tupper parameters $R$ and $a_2$
provide two  degrees of freedom, so there are many possible combinations
in the theory that give the correct value $\beta=1$.
In addition, if in the Page and Tupper
theory $a_1=1$, then from Eq.~\eqref{a12}
the theory has the correct
value for $\gamma$, $\gamma=1$.

\subsection{Ni's Lagrangian-based stratified theory}

This Ni's
theory \cite{ni1972} is a stratified scalar field theory of gravitation
and as such is an instance of the set of equations given
in Eqs.~\eqref{eq:poisson}, \eqref{eq:physmetric}, and 
\eqref{eq:partlag}.
This theory has a field Lagrangian density which yields the following field equation
\begin{align}
\label{eq:lagrfieldeq}
\frac{\partial}{\partial x^a} &\left[ \sqrt{-g} g^{ab} \frac{\partial \Phi}{\partial x^b} \right]
+ 2 \pi (-g)^{1/2} T^{ab} \frac{\partial g_{ab}}{\partial \Phi} \nonumber \\
& \qquad - \frac12 \frac{\partial \sqrt{-g} g^{ab}}{\partial \Phi} 
\frac{\partial \Phi}{\partial x^a} \frac{\partial \Phi}{\partial x^b} = 0 \\
&L = -mc \sqrt{e^{2\Phi/c^2}c^2 - e^{-2\Phi/c^2}v^2} \,.
\label{eq:nistratlag}
\end{align}
Using Eqs.~\eqref{eq:physmetric}, \eqref{eq:partlag}, and~\eqref{eq:nistratlag}, it is immediate to recover the physical metric $g_{ab}$ of this theory, and we get for the line element
$
ds^2 = - e^{2 \Phi/c^2}  dt^2 + e^{-2 \Phi/c^2} (dx^2+dy^2+dz^2)
$.

In a vacuum, this equation simplifies to
\begin{align}
\sqrt{-g} g^{ab} \frac{\partial^2 \Phi}{\partial x^a \partial x^b} + \frac12 \frac{\partial(\sqrt{-g} g^{ab})}{\partial \Phi} \frac{\partial \Phi}{\partial x^a}\frac{\partial \Phi}{\partial x^b} =0\,.
\end{align}
Using the fact that $\sqrt{-g}=e^{-2\Phi/c^2}$, we obtain 
\begin{align}
\sqrt{-g} g^{ab} = \textrm{diag} (-e^{-4\Phi},1,1,1)\,.
\end{align}
Therefore, although in general there is a self-interaction term, for the static case (which is the case we are interested in) that term vanishes and Eq.~\eqref{eq:lagrfieldeq} becomes $
\nabla^2 \Phi =0
$.
This implies that Eq.~\eqref{eq:lagrfieldeq} corresponds to $k=0$
in Eq.~\eqref{eq:poisson}.
The Lagrangian in Eq.~\eqref{eq:nistratlag} 
corresponds to $g(\Phi)= e^{2\Phi/c^2}$ and to $f(\Phi) = e^{-2\Phi/c^2}$ in  Eq.~\eqref{eq:partlag}.
Expanding to the post-Newtonian order, we get
for $g(\Phi)$ and $f(\Phi)$
\begin{align}
g(\Phi) =& 1 +  \frac{2\Phi}{c^2} + \frac{2 \Phi^2}{c^4} \,,\\
f(\Phi) =& 1 - \frac{2\Phi}{c^2} \,,
\end{align}
respectively.
Looking at Eqs.~\eqref{eq:g} and~\eqref{eq:f} 
this corresponds to
\begin{align}
A &= 1 \,, \\
B &= 1 \,, \\
C &= 1 \,.
\end{align}
Therefore, the PPN parameters $\beta$ and $\gamma$ defined
in
Eq.~\eqref{eq:PPN}, obeying the relation in 
Eq.~\eqref{eq:system}, 
are given by
\begin{align*}
\beta &= 1 \\
\gamma &= 1 \, .
\end{align*}
We can conclude that this theory correctly predicts the classical solar system tests.

\subsection{Ni's general conformally flat theory}

This Ni's
theory \cite{ni1972} is a conformally flat theory of gravitation
and as such is an instance of the set of equations given
in Eqs.~\eqref{eq:poisson}, \eqref{eq:physmetric}, and 
\eqref{eq:partlag}.
The field equation and particle Lagrangian in this theory are given by
\begin{align}
\label{eq:niconfield}
\square \Phi &= 4 \pi G K(\Phi) \rho \,,\\
\label{eq:niconlag}
L &= -m c \sqrt{e^{-2 F(\Phi)}  c^2 -  e^{-2 F(\Phi)} v^2} \,,
\end{align}
respectively, where $K(\Phi)=1 - p \Phi$ and $F(\Phi) = - \frac{\Phi}{c^2} + q \frac{\Phi^2}{c^4} + \dots$,
with $p$ and $q$ being two dimensionless constants.
Note that Nordström's theory \cite{nordstrom1913} is a particular case of this theory.
Using Eqs.~\eqref{eq:physmetric}, \eqref{eq:partlag}, and~\eqref{eq:niconlag}, it is immediate to recover the physical metric of this theory, and we get for the line element, i.e., the metric, 
$
g_{ab} = e^{-2 F(\Phi)} \gamma_{ab}
$,
where $\gamma_{ab}$ is a flat metric in Minkowski spacetime, as in Eq.~\eqref{eq:minkowski}.

The field equation Eq.~\eqref{eq:niconfield} of the
theory  corresponds to $k=0$
in  Eq.~\eqref{eq:poisson}.
The particle Lagrangian in Eq.~\eqref{eq:niconlag} 
corresponds to $g(\Phi)=e^{-2 F(\Phi)} $ and to $f(\Phi) = e^{-2 F(\Phi)}$ in  Eq.~\eqref{eq:partlag}.
Expanding to the post-Newtonian order, we get
for $g(\Phi)$ and $f(\Phi)$
\begin{align}
g(\Phi) =& 1 +  \frac{2\Phi}{c^2} + (1-q) \frac{2 \Phi^2}{c^4}\,,\\
f(\Phi) =& 1 + \frac{2\Phi}{c^2} \,,
\end{align}
respectively. Note, that $f(\Phi)$ also gets a $(1-q) \frac{2 \Phi^2}{c^4}$
in the expansion, but we do not need it in the calculations.
Looking at Eqs.~\eqref{eq:g} and~\eqref{eq:f} 
this corresponds to
\begin{align}
B &= 1-q \nonumber \,,\\
C &= -1 \nonumber \,.
\end{align}
Therefore, the PPN parameters defined
in
Eq.~\eqref{eq:PPN}, obeying the relation in 
Eq.~\eqref{eq:system}, 
are given by
\begin{align*}
\beta &= 1-q \\
\gamma &= -1\,.
\end{align*}
Therefore, since $\gamma\neq1$, this theory does not have the
correct post-Newtonian form that could explain solar system phenomena.

\subsection{A new conformal scalar theory of gravitation in flat spacetime}
\label{nct}

Following Freund and Nambu \cite{freund1968} (see also \cite{franklin2015} for the static case), one can build a general self-consistent relativistic scalar theory of gravitation in flat spacetime (see \ref{ap:theory}).

In this new theory, the vacuum field equation and particle Lagrangian are given by (see \ref{ap:theory})
\begin{align}
\label{eq:scalarfield}
\square \Phi &= - \frac{1}{1-\frac{2 \Phi}{c^2}} \frac{1}{c^2}(\nabla_c \Phi)(\nabla^c \Phi)\,,  \\
\label{eq:scalarlag}
L &= -mc^2 \sqrt{1-\frac{v^2}{c^2}} \left( 1 + \frac{\Phi h_2(\Phi)}{c^2} \right) \, ,
\end{align}
where $h_2$ is a dimensionless function satisfying $h_2(0)=1$.
Using Eqs.~\eqref{eq:physmetric}, \eqref{eq:partlag}, and~\eqref{eq:scalarlag}, it is immediate to recover the physical metric of this theory, and we get for the line element, i.e., the metric, 
$
g_{ab} =   \left( 1 + \frac{\Phi h_2(\Phi)}{c^2}  \right)^2 \gamma_{ab}
$,
where $\gamma_{ab}$ is a flat metric in Minkowski spacetime, as in Eq.~\eqref{eq:minkowski}.

The field equation Eq.~\eqref{eq:scalarfield} of the
theory corresponds to $k=1$ in Eq.~\eqref{eq:poisson}.
The particle Lagrangian in Eq.~\eqref{eq:scalarlag} 
corresponds to $g(\Phi)=f(\Phi)= \left( 1 + \Phi h_2(\Phi)/c^2 \right)^2$ in  Eq.~\eqref{eq:partlag}.
Expanding to the post-Newtonian order, and assuming that we can expand $h_2$ to first order in $\Phi$, we get
for $g(\Phi)$ and $f(\Phi)$
\begin{align}
g(\Phi) =& 1 +  \frac{2\Phi}{c^2} + \left(h_2'(0) +\frac12 \right) \frac{2 \Phi^2}{c^4}\,,\\
f(\Phi) =& 1 + \frac{2\Phi}{c^2} \,,
\end{align}
where $'$ means differentiation
with respect to $\Phi$.
Note again, that $f(\Phi)$ also gets a $\left(h_2'(0) +\frac12 \right) \frac{2 \Phi^2}{c^4}$
in the expansion, but we do not need it in the calculations.
Looking at Eqs.~\eqref{eq:g} and~\eqref{eq:f} 
this corresponds to
\begin{align*}
B &= h_2'(0) + \frac12 \\
C &= -1 \, , 
\end{align*}
Therefore, the PPN parameters defined
in
Eq.~\eqref{eq:PPN}, obeying the relation in 
Eq.~\eqref{eq:system}, 
are given by
\begin{align*}
\beta &= h_2'(0)\,, \\
\gamma &= -1\,.
\end{align*}
Even though we have a free parameter $h_2'(0)$
which could be adjusted to one, this theory always yields $\gamma=-1$, and therefore does not explain all the solar system tests. 
In order to solve this problem, one could think about relaxing equation Eq.~\eqref{eq:scalarfield} for self-coupling and allow a general $k$. This would only change the expression of $\beta$, which would become $\beta=h_2'(0) + \frac12 - \frac{k}{2}$; $\gamma$ would still be $-1$. This is why it is so challenging to build a relativistic scalar theory of gravitation that respects Lorentz symmetries.
Since this theory is very general, the only way to obtain the correct PPN parameters from a scalar field theory in flat spacetime would be to modify the field Lagrangian density by adding for instance a term proportional to $T^{ab} (\partial_a \Phi) (\partial_b \Phi)$.
The consequences of this modification cannot be straightforwardly derived with the formalism developed in this paper, since in this case the effective physical metric also depends on $\partial_a \Phi$ (that is $f$ and $g$ are functions of $\Phi$ and $\partial_a \Phi$) and here we assumed that $f$ and $g$ are only functions of $\Phi$, because this is what was used in the literature.
The extension of this model to account for derivative couplings is then left for future work.

\vspace{1cm}

\section{Conclusion}
\label{sec:conc}
In this paper, we presented a general stratified scalar field theory of gravitation in a Minkowski background. Then, we calculated two post-Newtonian parameters from three general parameters of the theory $B$, $C$ and $k$, concluding that it is perfectly possible for such a scalar theory to explain the four solar system tests. Finally, we used this general theory to rapidly compute the PPN parameters $\beta$ and $\gamma$ for a set of scalar theories of gravitation to verify if they agree with the experimental tests of gravitation in the solar system. 
Therefore, with this formalism, one can directly find those two PPN parameters only from the field equation and the particle Lagrangian of a given scalar theory of gravitation.
Although this is a very efficient method to calculate $\beta$ and $\gamma$ for a given theory, it does not allow one to compute the other PPN parameters. It would be interesting to generalize this approach to efficiently calculate the remaining PPN parameters for scalar theories and verify if it is possible for such a theory to explain every phenomenon predicted by general relativity.

The stratified theories that were analyzed (Page and Tupper's, and Ni's) yielded the correct PPN parameters relevant for solar system tests. 
One could wonder whether this indicates that they are valid theories, and the answer to that relies in analyzing the remaining PPN parameters. 
This analysis was done by Nordtvedt and Will \cite{nordtvedt1972} and Ni \cite{ni1972} and the conclusion was that stratified theories cannot account for Earth-tide measurements due to the motion of the solar system relative to the preferred frame (defined by the distant stars).

The conformal theories that were analyzed did not yield the correct $\gamma$ parameter even in very general cases. This motivates future work on the analysis of a relativistic scalar theory including a derivative coupling in the Lagrangian, of the type $T^{ab} (\partial_a \Phi) (\partial_b \Phi)$. Such a theory would not have preferred frame effects (it would respect Lorentz symmetries), so if it predicted the correct parameters $\beta$ and $\gamma$ it would not have the problem of Earth-tide measurements.

If such a scalar theory correctly predicts the outcome of every weak field gravity experiment, then we can only rule it out using strong gravity experiment results (e.g. LIGO, neutron star binaries, cosmology). Note also that a scalar theory of gravity is much simpler than general relativity, since it describes gravity with one function instead of ten. In such theories, unlike general relativity, it is generally possible to define a local gravitational energy-momentum tensor, which is always an attractive feature, and is still a problem in general relativity.

\section*{ACKNOWLEDGMENTS}

DPLB thanks a grant from
Funda\c c\~ao para a \\
Ci\^encia e Tecnologia
(FCT), Portugal, 
through project
No.~UID/FIS/00099/2013.
JPSL
thanks FCT for the grant No.~SFRH/BSAB/128455/2017, Coordena\c
c\~ao de Aperfei\c coamento do Pessoal de N\'\i vel Superior (CAPES),
Brazil, for support within the Programa CSF-PVE, Grant
No.~88887.068694/2014-00.

\appendix

\section{A new conformal scalar field theory of gravitation
in flat spacetime: Criteria and equations}
\label{ap:theory}
In this section, 
following the approach of Freund and Nambu \cite{freund1968} (see also Franklin \cite{franklin2015} for the static case) 
we build a new conformal
scalar theory of gravitation in flat Minkowski spacetime from a set of criteria,
see Sec.~\ref{nct}.
These criteria are
\begin{enumerate}
\item The spacetime metric is given by
\begin{align}
ds^2 = \gamma_{ab}\,dx^a\,dx^b\,,
\label{stm}
\end{align}
where $\gamma_{ab}$ is a flat metric not necessarily of Minkowski form,
i.e., not necessarily $\gamma_{ab}=\eta_{ab}={\rm diag}\,(-1,1,1,1)$.
\item In the Newtonian limit, the field Lagrangian density should be equal
to
\begin{align}
\mathcal{L}_{\rm Newt} = - \rho \Phi-\frac{1}{8 \pi G}|\nabla \Phi|^2  \,.
\end{align}
\item The general form of the Lagrangian is
\begin{align}
\label{eq:fieldlag}
\mathcal{L} = \mathcal{L_{\rm free}} +  \mathcal{L}_{\rm m} + \mathcal{L_{\rm int}}  \,,
\end{align}
where 
\begin{align}
\mathcal{L_{\rm free}}=\frac{h_1(\Phi)}{8 \pi G} (\partial_a \Phi) (\partial^a \Phi) \,,
\label{lfree}
\end{align}
with $h_1$ being a dimensionless function
to be determined which accounts for self-interaction and
satisfies $h_1(0)=-1$,
$\mathcal{L}_{\rm m}$ is the matter Lagrangian density, and
\begin{align}
\mathcal{L_{\rm int}}= \frac{\Phi}{c^2} h_2(\Phi)T_{\rm m} \,,
\end{align}
with $h_2$  a dimensionless free function satisfying $h_2(0)=1$, and
$T_{\rm m} $ is the trace of the matter energy-momentum tensor
$T_{\rm m}^{ab}$
defined as $
\sqrt{-\gamma}T_{\rm m}^{ab}=\frac {\delta \mathcal{L}_{\rm m}}{\delta \gamma_{ab}}$,
$\gamma$ being the determinant of $\gamma_{ab}$, and $\delta$ denotes
functional variation.
\item The energy-momentum tensor for the gravitational field is
given by Noether's expression
\begin{align}
\label{eq:emt}
T^{ab}_{\rm grav} = \frac{\partial \mathcal{L_{\rm
free}}}{\partial(\partial_a \Phi)} \partial^b \Phi - \gamma^{ab}
\mathcal{L_{\rm free}}\, ,
\end{align}
where $\mathcal{L_{\rm free}}$ is the free field Lagrangian of Eq.~\eqref{lfree}.
\item  The exact free field equation should be of the form
\begin{align}
\label{eq:fieldeq}
\square \Phi =\kappa T_{\rm grav}\, ,
\end{align}
in order to account explicitly for the self interaction of the field, where
$T_{\rm grav}$ is the trace of 
$T^{ab}_{\rm grav} $ and 
$\alpha$ is a coupling constant to be determined.
\item  
In a static vacuum, the field equation (\ref{eq:fieldeq}) should simplify to
\begin{align}
\label{eq:selfenergy}
\nabla^2 \Phi = -\frac{|\nabla \Phi|^2}{c^2} \,.
\end{align}

\end{enumerate}

With these requirements in hand, we begin by calculating the expression
for $T^{ab}_{\rm grav}$ using
Eqs.~\eqref{lfree} and \eqref{eq:emt}. We then obtain
\begin{align}
T^{ab}_{\rm grav} = \frac{h_1(\Phi)}{4 \pi G} \left[
(\partial^a \Phi) (\partial^b \Phi) - \frac12
\gamma^{ab} (\partial_c \Phi) (\partial^c \Phi) \right]\,.
\end{align}
Calculating the trace yields
\begin{align}
\label{eq:traceemt}
T_{\rm grav} = -\frac{h_1(\Phi)}{4 \pi G} (\partial_c \Phi) (\partial^c \Phi)\,.
\end{align}
First we fix the constant of proportionality $\kappa$ in
Eq.~\eqref{eq:fieldeq}. In the static, Newtonian limit,
where $h_1(\Phi)=-1$, we have $T_{\rm grav} = \frac{1}{4 \pi G} |\nabla \Phi|^2$.
Therefore, in order to account for Eq.~\eqref{eq:selfenergy}
one has $\kappa=-\frac{4 \pi G}{c^2}$ and the field equation in vacuum is
\begin{align}
\label{eq:fieldequation}
\square \Phi = -\frac{4 \pi G}{c^2} T_{\rm grav}\,.
\end{align}
Second we determine $h_1$.
The Euler-Lagrange equation give for the free Lagrangian Eq.~\eqref{lfree},
\begin{align}
\label{eq:elequation}
\square \Phi = - \frac{h_1'(\Phi)}{2 h_1(\Phi)} (\partial_c \Phi) (\partial^c \Phi)\,. 
\end{align}
Replacing Eqs.~\eqref{eq:traceemt} and \eqref{eq:elequation}
in Eq.~\eqref{eq:fieldequation} yields the following differential equation for $h_1$,
$
h_1' = -\frac{2}{c^2} h_1^2 \,,
$
which upon integration gives, considering that $h_1(0)=-1$,
\begin{align}
h_1(\Phi) = - \frac{1}{1-\frac{2 \Phi}{c^2}} \,.
\label{h1}
\end{align}
Thus, using Eq.~\eqref{h1} in 
Eq.~\eqref{eq:traceemt} together with \eqref{eq:fieldequation}, or
directly in \eqref{eq:elequation}, we obtain
the field equation for the gravitational field $\Phi$ in vacuum,
\begin{align}
\square \Phi
=-\frac{1}{c^2\left(1-\frac{2 \Phi}{c^2}\right)}
(\partial_a \Phi)
(\partial^a \Phi) \,.
\end{align}
The full field equation, i.e., the equation derived taking into account 
$\mathcal{L_{\rm free}}$ and $\mathcal{L_{\rm int}}$
in Eq.~\eqref{eq:fieldlag} is then
\begin{multline}
\square \Phi
=-\frac{1}{c^2\left(1-\frac{2 \Phi}{c^2}\right)}
(\partial_a \Phi)
(\partial^a \Phi) - \\
\frac{4\pi G}{c^2}\left(1-\frac{2 \Phi}{c^2}\right)
(h_2+\Phi h_2')T_{\rm m}\,.
\label{eq:fieldlag2}
\end{multline}

Finally, we want to find an expression for the matter
Lagrangian from $\mathcal{L}_{\rm m}$
and $\mathcal{L_{\rm int}}$ in  Eq.~\eqref{eq:fieldlag}.
Since we want to compute how particles behave in the theory
our matter is represented by a point particle.
To simplify the analysis
we use Minkowski coordinates, i.e.,
$\gamma_{ab}=\eta_{ab}$, $\eta_{ab}={\rm diag}(-1,1,1,1)$.
In this case the matter Lagrangian density is the Lagrangian
density for a point particle
\begin{align}
\mathcal{L}_{\rm m} &= \rho_0 \eta_{ab}u^a u^b \nonumber \\
&= m \delta^3(\boldsymbol{x}-\boldsymbol{x_0})
\sqrt{1-\frac{v^2}{c^2}}\eta_{ab}u^a u^b\, ,
\label{a1}
\end{align}
where $\rho_0$ is the scalar proper mass density,
$\boldsymbol{x}$ represents spatial coordinates
and $\boldsymbol{x_0}$ the spatial position of the
particle,
and $u^a$ is the particle's four-velocity with respect to the
metric $\eta_{ab}$.
The matter energy-momentum tensor is the 
energy-momentum tensor for a point particle
determined from  $
T_{\rm m}^{ab}=\frac {\delta \mathcal{L}_{\rm m}}{\delta \eta_{ab}}$
\cite{Doughtylagrangian1990}
\begin{align}
T^{ab}_{\rm m} &= \rho_0 u^a u^b \nonumber \\
&= m \delta^3(\boldsymbol{x}-\boldsymbol{x_0}) \sqrt{1-\frac{v^2}{c^2}} u^a u^b \, .
\label{a2}
\end{align}
This yields the trace
\begin{align}
T_{\rm m} = - mc^2 \delta^3(\boldsymbol{x}-\boldsymbol{x_0}) \sqrt{1-\frac{v^2}{c^2}} \,.
\end{align}
Thus, using Eqs.~\eqref{a1} and \eqref{a2} for the sum
$\mathcal{L}_{\rm m} +\mathcal{L}_{\rm int}$ that appears in Eq.~\eqref{eq:fieldlag}
we have
$\mathcal{L}_{\rm m} +\mathcal{L}_{\rm int}=
 -m c^2\delta^3(\boldsymbol{x}-\boldsymbol{x_0}) \sqrt{1-\frac{v^2}{c^2}}
 \left(1 + \frac{\Phi}{c^2} h_2(\Phi)\right)
$, where we have used
$\eta_{ab}u^a u^b=-c^2$.
Integrating over all space, we get the matter plus the interaction Lagrangian
for the particle
which we simply call the particle Lagrangian $L$,
$L = \int d^3x (\mathcal{L}_{\rm m} +\mathcal{L}_{\rm int})$,
i.e., 
\begin{align}
L = -mc^2 \sqrt{1-\frac{v^2}{c^2}} \left( 1 + \frac{\Phi h_2(\Phi)}{c^2} \right) \,.
\label{pl1}
\end{align}

The field equation for the gravitational field~\eqref{eq:fieldlag2}
together with the particle Lagrangian \eqref{pl1} are the equations of
this theory and this is all we need to know in order to calculate the
trajectory of particles.

One could make the theory even more interesting by modifying the field
Lagrangian density through the addition of a term proportional to
$T^{ab} (\partial_a \Phi) (\partial_b \Phi)$.

\end{document}